\documentclass[11pt,cite]{article}
\usepackage{latexsym}

\usepackage[latin1]{inputenc}

\usepackage{graphicx}
\usepackage{color,pst-plot}

\usepackage{amsmath}
\usepackage {amsfonts,amssymb,amstext}

\newcommand{\lbl}[1]{\label{eq:#1}}
\newcommand{ \rf}[1]{(\ref{eq:#1})}

\newcommand{\be}{\begin{equation}}
\newcommand{\ee}{\end{equation}}
\newcommand{\bea}{\begin{eqnarray}}
\newcommand{\eea}{\end{eqnarray}}
\newcommand{\setl}{\setlength\arraycolsep{2pt}}

\newcommand{\noi}{\noindent}
\newcommand{\nn}{\nonumber}
\newcommand{\ra}{\rightarrow}

\newcommand{\cF}{{\cal F}}

\newcommand{\cM}{{\cal M}}

\newcommand{\cO}{{\cal O}}

\newcommand{\Imm}{\mbox{\rm Im}}
\newcommand{\Ree}{\mbox{\rm Re}}

\newcommand{\annd}{\mbox{\rm and}}

\newcommand{\als}{\alpha_{\mbox{\rm {\scriptsize s}}}}

\input epsf

\begin{document}

\begin{titlepage}

\begin{flushright}
\today  \\
\end{flushright}

\vspace*{0.2cm}
\begin{center}

{\Large{\bf Moment Analysis of Hadronic Vacuum Polarization}}\\\vspace*{0.25cm} 
{\large{\bf Proposal for a lattice QCD evaluation of $g_{\mu}-2$}}
\\[2 cm]

 {\bf Eduardo de Rafael}~$^{a}$
\\[1cm]

$^a$  {\it Aix-Marseille Universit\'e, CNRS, CPT, UMR 7332, 13288 Marseille, France}
    
\end{center} 

\vspace*{3.0cm}   

\begin{abstract}
I suggest a new approach to the determination of the hadronic vacuum polarization (HVP) contribution to the anomalous magnetic moment of the muon $a_{\mu}^{\rm HVP}$ in lattice QCD. It is  based on properties of the Mellin transform of the hadronic spectral function and their relation to the HVP self energy in the Euclidean. I show how $a_{\mu}^{\rm HVP}$ is very well approximated by a few moments associated to this Mellin transform and how these moments can be evaluated in lattice QCD, providing thus a series of tests when compared with the corresponding determinations using experimental data. 
\end{abstract}

\end{titlepage}

\noi
{\bf 1.} The hadronic vacuum polarization (HVP)contribution to the  anomalous magnetic moment of the muon, when expressed in terms of the HVP self energy $\Pi(Q^2)$ in the Euclidean ($Q^2 \ge 0$), is given by the Feynman parametric integral~\cite{LdeR69,LPdeR72}:
\be\lbl{eq:LdeR}
a_{\mu}^{\rm HVP} =  
\frac{\alpha}{\pi}\int_{0}^{1} dx (1-x)
\left[-\Pi\left(Q^2 \equiv \frac{x^2}{1-x}m_{\mu}^2 \right) \right]\,.
\ee
The on-shell renormalized function $\Pi(Q^2)$ obeys the dispersion relation 
\be\lbl{eq:disprel}
\Pi(Q^2) =\int_{4 m_{\pi}^2}^{\infty}\frac{dt}{t}\,
\frac{-Q^2}{t+Q^2}\frac{1}{\pi}\Imm\Pi(t)\,,
\ee
and the hadronic spectral function $\frac{1}{\pi}\Imm\Pi(t)$ is directly accessible to experiment via the one photon $e^+ e^-$ annihilation cross section into hadrons ($m_e \ra 0$):
\be\lbl{eq:sigmaee}
\sigma(t)=\frac{4\pi^2 \alpha}{t}\frac{1}{\pi}\Imm\Pi(t)\,.
\ee
Inserting Eqs.~\rf{eq:disprel} and \rf{eq:sigmaee} in the r.h.s. of Eq.~\rf{eq:LdeR} reproduces the standard representation used in all the phenomenological evaluations of $a_{\mu}^{\rm HVP}$~\footnote{For a recent review article on the muon $g-2$ experiments and theoretical evaluations see e.g. ref.~\cite{MdeRRS12}.}.

In lattice QCD evaluations of $a_{\mu}^{\rm HVP} $~\cite{Blum03}, it seems convenient to trade the Feynman $x$-parameter in Eq.~\rf{eq:LdeR}  by the Euclidean $Q^2$ momenta
with the results $\left(\omega=\frac{Q^2}{m_{\mu}^2} \right)$:

{\setl
\bea \lbl{eq:eucint0}
a_{\mu}^{\rm HVP} & = & \frac{\alpha}{\pi}\int_0^\infty \frac{d\omega}{\omega}\sqrt{\frac{\omega}{4+\omega}}\left(\frac{\sqrt{4+\omega}-\sqrt{\omega}}{\sqrt{4+\omega}+\sqrt{\omega}} \right)^2\left[-\Pi\left(\omega m_{\mu}^2 \right) \right]\,, \\ \lbl{eq:eucint} 
 & = & \frac{\alpha}{\pi}\int_0^\infty d\omega\ \frac{1}{4}
\left[\left(2+\omega \right)\left(2+\omega-\sqrt{\omega}\sqrt{4+\omega}\right)-2\right] \left(-\frac{d}{d\omega}\Pi\left(\omega m_{\mu}^2 \right) \right)\,. 
\eea}

\noi
Lattice QCD determinations of $\Pi\left(\omega m_{\mu}^2 \right)$ and/or $\frac{d}{d\omega}\Pi\left(\omega m_{\mu}^2 \right)$ at a sufficiently high enough number of values of $\omega$ could, in principle, provide an evaluation of these integrals with an accuracy perhaps comparable or eventually even better than the phenomenological determinations which use experimental data. At present, however, this is certainly not the case and so far the lattice determinations have to be complemented either by functional forms inspired by models or by other methods like Pad\'e approximants~\cite{ABGP12,FJMW13,HPQCD14},  which extrapolate the behaviour of $\Pi\left(\omega m_{\mu}^2 \right)$ and/or $\frac{d}{d\omega}\left[\Pi\left(\omega m_{\mu}^2 \right) \right]$  to the regions not covered by the lattice data, in particular the region at low $\omega$  which  is heavily weighted by the kernels in Eqs.~\rf{eq:eucint0} and/or \rf{eq:eucint} and, therefore, introduces large uncertainties. 

\vspace*{0.25cm} 

\noi
{\bf 2.} I suggest making a new type of evaluation of $a_{\mu}^{\rm HVP}$ which I call {\it the moment analysis}. It is based on the observation that the function $\frac{d}{d\omega}\Pi\left(\omega m_{\mu}^2 \right)$  has the Mellin--Barnes integral representation~\footnote{For an application of this technique to the evaluation of QED contributions to $g_{\mu}-2$ see ref.~\cite{AdeRG08}.}
\be\lbl{eq:BranesPi}
-\frac{d}{d\omega}\Pi\left(\omega m_{\mu}^2 \right)=\int_{4 m_{\pi}^2}^\infty\frac{dt}{t} \frac{m_{\mu}^2}{t}\frac{1}{2 \pi i}\int_{c-i\infty}^{c+i\infty}ds \left(\frac{\omega m_{\mu}^2}{t}\right)^{-s}\Gamma(s)\Gamma(2-s)
\frac{1}{\pi}\Imm\Pi(t)\,,
\ee
which follows from  the dispersion relation in Eq.~\rf{eq:disprel} and the identity:
\be
\frac{1}{(1+A)^2}=\frac{1}{2 \pi i}\int_{c-i\infty}^{c+i\infty}ds \left(A\right)^{-s}\ \Gamma(s)\Gamma(2-s)\,.
\ee
Inserting this representation in the r.h.s. of Eq.~\rf{eq:eucint} and performing the integration over $\omega$ results in a useful Mellin-Barnes representation for $a_{\mu}^{\rm HVP}$:
\be\lbl{eq:MBrepex}
a_{\mu}^{\rm HVP}
=  \left(\frac{\alpha}{\pi}\right) \frac{1}{2\pi i}\int\limits_{c-i\infty}^{c+i\infty}ds\ \cF(s)\ \cM(s)\,,
\ee
where $\cF(s)$ is a known function:
\be
\cF(s)= -\Gamma(3-2s)\Gamma(-3+s)\Gamma(1+s)\,,
\ee
and $\cM(s)$ the Mellin transform of the hadronic spectral function
\be\lbl{eq:mellinsp}
\cM(s)=\int_{4 m_{\pi}^2}^\infty\frac{dt}{t}\left(\frac{m_{\mu}^2}{t}\right)^{1-s}\frac{1}{\pi}\Imm\Pi(t)\,.
\ee
The Mellin transform in QCD is finite for $s<1$ and singular at $s=1$ with a residue fixed by perturbative QCD (pQCD). At leading order, with  three light active quarks $u$, $d$ and $s$, and with neglect of $\als$ corrections (which in any case can be included if necessary):
\be\lbl{eq:pqcd}
\cM_{\rm pQCD}(s)\underset{{s\ra\ 1}}{\thicksim} \left(\frac{\alpha}{\pi}\right)\left(\frac{2}{3}\right)N_c\  \frac{1}{3}\ \frac{1}{1-s}\,.
\ee 

The reason why the representation in Eq.~\rf{eq:MBrepex} is useful is that one can easily extract from it the asymptotic expansion for $\frac{m_{\mu}^2}{t}< 1$. This expansion is governed by the residues of the singularities of the integrand at the left of the {\it fundamental strip} (defined in our case by $\Ree~c \in\ ]0,+1[~$~\cite{FGD95}). The singularities in question are a single leading pole at $s=0$ and single and double poles at $s=-n$ with $n=1,2,...$.  The residues of these singularities are given by  the Mellin transform in Eq.~\rf{eq:mellinsp} at the values
\be
\cM(-n)=\int\limits_{4m_{\pi}^2}^{\infty}\frac{dt}{t}\left(\frac{m_{\mu}^2}{t} \right)^{1+n}
\frac{1}{\pi}\Imm\Pi(t)\,, \quad n=0,1,2,\dots \,,
\ee
and, because of the double poles of $\cF(s)$ at $s=-1,-2,\dots$, also by the first derivative of the Mellin transfom
\be
{\tilde{\cM}(s)}=-\frac{d}{ds}\cM(s)=\int_{4 m_{\pi}^2}^\infty \frac{dt}{t}\left(\frac{m_{\mu}^2}{t} \right)^{1-s}\log\frac{m_{\mu}^2}{t}\ \frac{1}{\pi}\Imm\Pi(t)
\ee
at the values:
\be\lbl{eq:logwe}
\tilde{\cM}(-n)=\int_{4 m_{\pi}^2}^\infty\frac{dt}{t}\ \left(\frac{m_{\mu}^2}{t}\right)^{1+n}\ \log\frac{m_{\mu}^2}{t}\ \frac{1}{\pi}\Imm\Pi(t)\,,\quad n=1,2,3,\cdots\,.
\ee

The explicit evaluation of $a_{\mu}^{\rm HVP}$ in terms of the moments $\cM(-n)$ and $\tilde{\cM}(-n)$ proceeds as follows.  The {\it singular expansion} of $\cF(s)$ at the l.h.s. of the {\it fundamental strip} is
\begin{equation}
\cF(s)
\asymp 
\frac{1}{3}\frac{1}{s}-\frac{1}{(s+1)^2}+\frac{25}{12}\frac{1}{s+1}-\frac{6}{(s+2)^2}+\frac{97}{10}\frac{1}{s+2}-\frac{28}{(s+3)^2}+\frac{208}{5}\frac{1}{s+3}+\cdots\,,
\end{equation}
and from this, the expansion of $a_{\mu}^{\rm HVP}$ in terms of successive moment approximants can be easily obtained with the result

{\setl
\bea\lbl{eq:MAn}
a_{\mu}^{\rm HVP} &  = & \left(\frac{\alpha}{\pi}\right) \left\{ \frac{1}{3}\cM(0) +\frac{25}{12} \cM(-1)+
\tilde{\cM}(-1)\right. \nn \\
 & + &  
\frac{97}{10}  \cM(-2)+6
\tilde{\cM}(-2) \nn \\
 & + & \left.
\frac{208}{5} \cM(-3)+28
\tilde{\cM}(-3) 
+\cO\left[\tilde{\cM}(-4)\right]\right\}\,.
\eea}

\noi
The  $\cM$ moments give positive contributions while the  $\tilde{\cM}$ moments give negative contributions which in absolute value are larger than those of the corresponding $\cM$ moments. Numerically, because of the $\rho$--dominance of the hadronic spectral function and the fact that $\frac{m_{\mu}^2}{M_{\rho}^2}\simeq 1.9\times 10^{-2}$ is a small number,
only a few moments are necessary to get an accurate evaluation, a fact which we next illustrate within the framework of a realistic phenomenological toy model.

\vspace*{0.25cm}

\noi
{\bf 3.}
The model in question is the one described in ref.~\cite{BM11}~\footnote{with some modifications kindly contributed by Laurent Lellouch.} 
The evaluation of $a_{\mu}^{\rm HVP}$ in this model gives:
\be
a_{\mu}^{\rm HVP}(\rm phen.~model)=6.936\times 10^{-8}\,,
\ee
in agreement with the determination from $e^+ e^-$ data~\cite{Detal11}
\be\lbl{eq:exp}
a_{\mu}^{\rm HVP}(e^+ e^-)=(6.923\pm 0.042)\times 10^{-8}\,.
\ee
\begin{figure}[h]
\hspace*{-0.8cm}\includegraphics[width=0.53\textwidth]{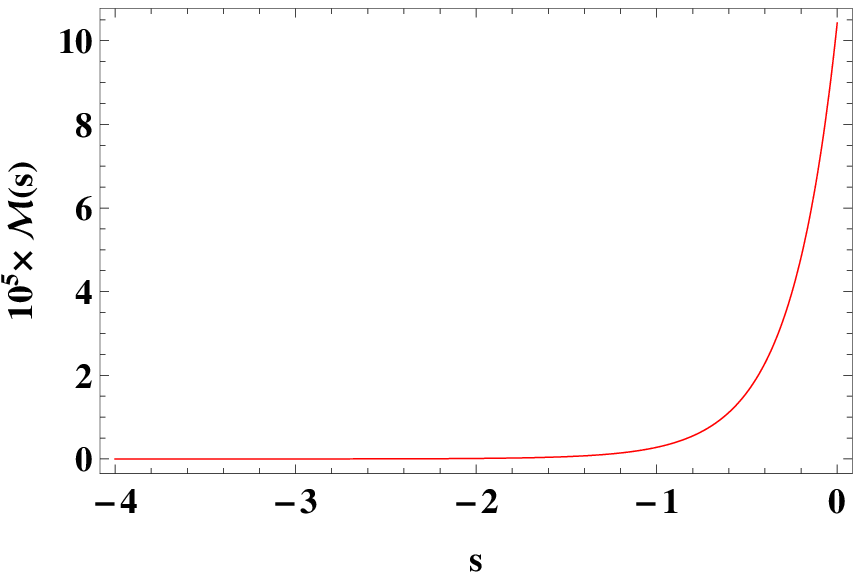} 
\includegraphics[width=0.55\textwidth]{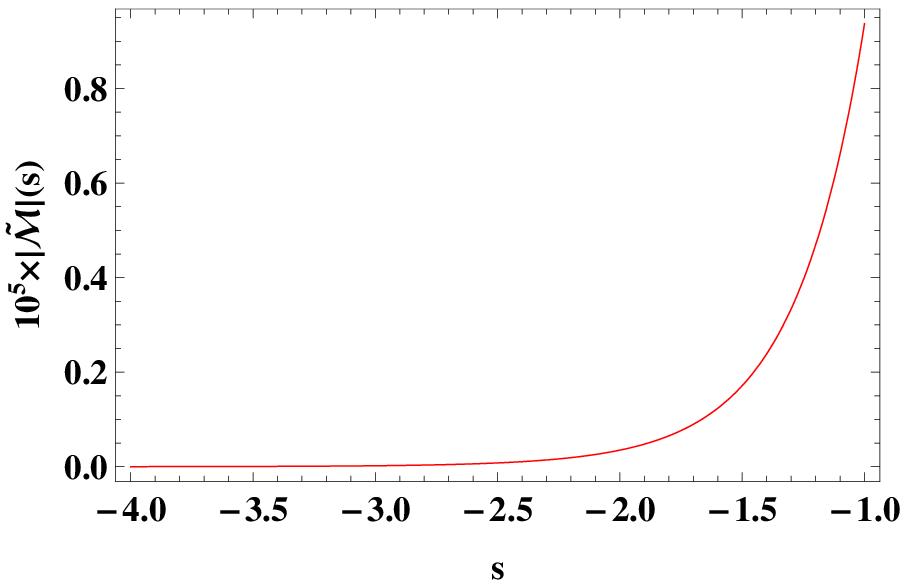}

\begin{center}
{\bf Fig.~1~~{\it  The Mellin Transforms $\cM(s)$ and $\vert{\tilde\cM}\vert(s)$ in the toy model of ref.~\cite{BM11}.}}
\end{center}
\end{figure}
\noi
The shape of the Mellin transform  and its derivative  in this model  are shown in Fig.~1. As seen in these figures these Mellin transforms are sharply decreasing functions for negative $s$-values, and very smooth compared to the shape of the hadronic spectral function.  
The results in this model, corresponding to the successive moment approximants in Eq.~\rf{eq:MAn}, are:
\be\lbl{eq:zero}
\left(\frac{\alpha}{\pi}\right)  \frac{1}{3}\cM(0)=8.071\times 10^{-8}\,,
\ee
\be\lbl{eq:zmo}
\left(\frac{\alpha}{\pi}\right) \left[ \frac{1}{3}\cM(0) +\frac{25}{12} \cM(-1)+
\tilde{\cM}(-1) \right]=7.240\times 10^{-8}\,,
\ee
\be\lbl{eq:zmt}
\left(\frac{\alpha}{\pi}\right) \left[ \frac{1}{3}\cM(0) +\frac{25}{12} \cM(-1)+
\tilde{\cM}(-1) +\frac{97}{10}  \cM(-2)+6
\tilde{\cM}(-2) \right]=7.022\times 10^{-8}\,.
\ee
The first approximation exceeds the phenomenological result by less than $16\%$, the second approximation by $4\%$, and the third approximation by $1\%$. In fact the fourth approximation results in an overestimate by only $0.4\%$ which is already of the same order of accuracy as the present experimental determination in Eq.~\rf{eq:exp} ($0.6\%$). This gives an idea of how many moments should be  determined in order to be competitive with the determinations of $a_{\mu}^{\rm HVP} $ which use experimental data.

\vspace*{0.25cm}

\noi
{\bf 4.}
The leading term in  the moment expansion in Eq.~\rf{eq:MAn} coincides with a rigorous upper bound discussed a long time ago~\cite{BdeR69}:
\be\lbl{eq:BdeR}
a_{\mu}^{\rm HVP} < \left(\frac{\alpha}{\pi}\right)\frac{1}{3}\int_{4 m_{\pi}^2}^\infty\frac{dt}{t}\ \frac{m_{\mu}^2}{t}\ \frac{1}{\pi}\Imm\Pi(t)
=\left(\frac{\alpha}{\pi}\right)\frac{1}{3}
\left(-m_{\mu}^2\frac{d}{d Q^2}\Pi(Q^2)\right)_{Q^2 =0}\,.
\ee
It overestimates the phenomenological determination of $a_{\mu}^{\rm HVP}$ by less than $20\%$ (which is not bad for a rigorous bound) but what is more important here is the fact that it provides an excellent first check between lattice QCD evaluations and phenomenological determinations. Indeed, the second expression in the r.h.s. is the slope of $\Pi\left( Q^2\right)$ at the origin, a quantity which can be evaluated in lattice QCD and the accuracy of its determination compared to the one of the  phenomenological  determination of the first moment of the spectral function, the first term in the r.h.s. It is difficult to imagine that, unless lattice QCD does better than phenomenology in this simple case, it will ever reach a competitive accuracy of the full determination of $a_{\mu}^{\rm HVP}$. 

In general, the moments $\cM(-n)$ correspond to successive derivatives of the HVP self-energy $\Pi(Q^2)$ at the origin: for $n=0,1,2,\dots\,,$
\be\lbl{eq:momeucl}
\cM(-n)=\int\limits_{4m_{\pi}^2}^{\infty}\frac{dt}{t}\left(\frac{m_{\mu}^2}{t} \right)^{1+n}
\frac{1}{\pi}\Imm\Pi(t)=
\frac{(-1)^{n+1} }{(n+1)!}( m_{\mu}^2 )^{n+1} \left(\frac{\partial^{n+1}}{(\partial Q^2 )^{n+1}}\Pi(Q^2)\right)_{Q^2 =0}\,,
\ee
providing thus  a series of further tests of lattice QCD results to be compared with the moments obtained from experimental or phenomenological input of the hadronic spectral function.

The determination of the log weighted moments ${\tilde\cM}(-n)$ in Eq.~\rf{eq:logwe} in terms of the HVP self-energy function $\Pi(Q^2)$ is more delicate. It requires the evaluation of integrals of the type 
\be\lbl{eq:sigmaints}
\Sigma(-n)\equiv \int_{4m_{\pi}^2}^{\infty}d Q^2 \left(\frac{m_{\mu}^2}{Q^2}\right)^{n+1} \ \left(-\frac{\Pi(Q^2)}{Q^2}\right)\,,\quad n=1,2,3\cdots\,. 
\ee
To see this in detail 
let me discuss the evaluation of the first two moments  ${\tilde\cM}(-1)$ and ${\tilde\cM}(-2)$. (The generalization to the evaluations of higher $\tilde{\cM}$ moments is straightforward.)

One first observes that
\be\lbl{eq:fourpilog}
\tilde{\cM}(-n)=-\log\frac{4 m_{\pi}^2}{m_{\mu}^2}\cM(-n)+
\int_{4 m_{\pi}^2}^\infty \frac{dt}{t}\left(\frac{m_{\mu}^2}{t} \right)^n \log\frac{4 m_{\pi}^2}{t}\frac{1}{\pi}\Imm \Pi(t)\,,
\ee
which translates the problem to the evaluation of  $\log\frac{4 m_{\pi}^2}{t}$ weighted moments, which are smaller in magnitude.  
Using the dispersion relation in Eq.~\rf{eq:disprel} one can then show that

{\setl
\bea\lbl{eq:sig1}
\lefteqn{\hspace*{-1.5cm}\Sigma(-1)\equiv
\int_{4m_{\pi}^2}^{\infty}d Q^2\left(\frac{m_{\mu}^2}{Q^2}\right)^2 \left(-\frac{\Pi(Q^2)}{Q^2}\right)   =  \int\limits_{4m_{\pi}^2}^{\infty}\frac{dt}{t}\left(\frac{m_{\mu}^2}{t}\right)^2
\ \log\frac{4m_{\pi}^2}{t}\ \frac{1}{\pi}\Imm\Pi(t)} \nn \\ & &  
+\frac{m_{\mu}^2}{4m_{\pi}^2}\cM(0)-
\int\limits_{4m_{\pi}^2}^{\infty}\frac{dt}{t}\left(\frac{m_{\mu}^2}{t}\right)^2
\ \log\left(1+\frac{4m_{\pi}^2}{t}\right)\ \frac{1}{\pi}\Imm\Pi(t)\,,
\eea}

\noi
where the wanted $\log\frac{4 m_{\pi}^2}{t}$ weighted moment is the first term in the r.h.s. and the rest of the contributions can be expressed in terms of normal $\cM$ moments. From Eqs.~\rf{eq:fourpilog} and \rf{eq:sig1} there follows then that:
\be\lbl{eq:sigm1}
\tilde{\cM}(-1)=-\log\frac{4 m_{\pi}^2}{m_{\mu}^2}\cM(-1)+\Sigma(-1)  -\frac{m_{\mu}^2}{4m_{\pi}^2}\cM(0)+\frac{4 m_{\pi}^2}{m_{\mu}^2}\cM(-2)+\cdots\,.
\ee

\noi
Integrating next $\Pi(Q^2)$ with an extra power of $\frac{m_{\mu}^2}{Q^2}$  gives the new relation

{\setl
\bea\lbl{eq:sig2}
\lefteqn{\Sigma(-2)\equiv 
\int_{4m_{\pi}^2}^{\infty}d Q^2\left(\frac{m_{\mu}^2}{Q^2}\right)^3 \left(-\frac{\Pi(Q^2)}{Q^2}\right)   =  -\int\limits_{4m_{\pi}^2}^{\infty}\frac{dt}{t}\left(\frac{m_{\mu}^2}{t}\right)^3
\ \log\frac{4m_{\pi}^2}{t}\ \frac{1}{\pi}\Imm\Pi(t)} \nn \\ & &  
+\frac{1}{2}\left(\frac{m_{\mu}^2}{4m_{\pi}^2}\right)^2 \cM(0)-\frac{m_{\mu}^2}{4m_{\pi}^2}\cM(-1)+
\int\limits_{4m_{\pi}^2}^{\infty}\frac{dt}{t}\left(\frac{m_{\mu}^2}{t}\right)^3
\ \log\left(1+\frac{4m_{\pi}^2}{t}\right)\ \frac{1}{\pi}\Imm\Pi(t)\,,
\eea}

\noi
and, from this and Eq.~\rf{eq:fourpilog}:
\be\lbl{eq:sigm2}
\tilde{\cM}(-2)=-\log\frac{4 m_{\pi}^2}{m_{\mu}^2}\cM(-2)-\Sigma(-2) +\frac{1}{2}\left(\frac{m_{\mu}^2}{4m_{\pi}^2}\right)^2 \cM(0)-\frac{m_{\mu}^2}{4 m_{\pi}^2}\cM(-1)+
\frac{4 m_{\pi}^2}{m_{\mu}^2}\cM(-3)+\cdots\,.
\ee

\noi

From the relations above one concludes that the quantities to be evaluated in  lattice QCD are, therefore, the Euclidean moment integrals in Eq.~\rf{eq:sigmaints}.
Contrary to the direct evaluation of $a_{\mu}^{\rm HVP}$ in Eqs.~\rf{eq:eucint0} and/or \rf{eq:eucint}, the   moments $\Sigma(-1)$, $\Sigma(-2)$, ... are not weighted by a heavily peaked kernel at small $Q^2$ values and, furthermore, the threshold of integration is at the rather large value $Q^2=4m_{\pi}^2$ instead of zero, which makes them rather accessible to a lattice QCD evaluation. The determination of these integral moments and their comparison with the corresponding phenomenological expressions in terms of the hadronic spectral function given above, can provide valuable further tests. 
\vspace*{0.25cm}

\noi
{\bf 5.}
One can finally proceed to the evaluation of  successive approximations to $a_{\mu}^{\rm HVP}$ by replacing the expansion in terms of the $\cM$ moments and log weighted ${\tilde\cM}$ moments in Eqs.~\rf{eq:zmo} and \rf{eq:zmt} by the corresponding one in terms of the ordinary moments   $\cM$ and the integral  $\Sigma$ moments in Eq.~\rf{eq:sigmaints} discussed above. This leads to the following results:

\begin{itemize}
	\item {\bf 1st Approximation}
	
\be
 \left(\frac{\alpha}{\pi}\right)  \frac{1}{3}\cM(0)=8.071\times  10^{-8}\,.
\ee	

\item {\bf 2nd Approximation}

\be
 \hspace*{-1cm}\left(\frac{\alpha}{\pi}\right) \left\{ \left(\frac{1}{3}-\frac{m_{\mu}^2}{4 m_{\pi}^2}\right)\cM(0)+\left(\frac{25}{12}-\log\frac{4 m_{\pi}^2}{m_{\mu}^2} \right)\cM(-1)+\Sigma(-1)+\frac{4m_{\pi}^2}{m_{\mu}^2}\cM(-2)\right\}
\ee
\be
\hspace*{-9cm}=7.265(34) \times 10^{-8}\nn \,.
\ee

\item {\bf 3rd Approximation}

{\setl
\bea
\left(\frac{\alpha}{\pi}\right) \left\{ \left(\frac{1}{3}-\frac{m_{\mu}^2}{4 m_{\pi}^2}+3\left(\frac{m_{\mu}^2}{4 m_{\pi}^2}\right)^2\right)\cM(0)+\left(\frac{25}{12}-\log\frac{4 m_{\pi}^2}{m_{\mu}^2}-6\frac{m_{\mu}^2}{4 m_{\pi}^2} \right)\cM(-1)\right.\nn \\
& & \hspace*{-15cm}
\left. +\left(\frac{97}{10}-6\log\frac{4m_{\pi}^2}{m_{\mu}^2}+\frac{4 m_{\pi}^2}{m_{\mu}^2} \right)\cM(-2)+\Sigma(-1)-6\Sigma(-2)+\frac{4m_{\pi}^2}{m_{\mu}^2}\left(6-\frac{1}{2} \frac{4m_{\pi}^2}{m_{\mu}^2}\right)\cM(-3) \right\} \\ & & \hspace*{-13cm}
 =7.027(6)\times 10^{-8}\,.\nn
\eea}

\noi

\end{itemize}
\noi
The numerical results are those obtained in the phenomenological toy model  described above with
the quoted uncertainties in the second  and third approximations corresponding to the size of the first contributions which have not been retained in the expansions of the $\log\left(1+ \frac{4m_{\pi}^2}{t}\right)$ terms in Eqs.~\rf{eq:sig1} and \rf{eq:sig2}.

The relevant quantities to be determined in lattice QCD in order to construct the  three successive approximations above are therefore:
\be
\underbrace{\cM(0)}_{10.424}\,;\quad\underbrace{\Sigma(-1)}_{1.223}\,,\quad\underbrace{\cM(-1)}_{0.278}\,;\quad\underbrace{\Sigma(-2)}_{0.113}\,, \quad\underbrace{\cM(-2)}_{0.012}\,\quad\annd\quad\underbrace{\cM(-3)}_{0.001}\,,
\ee
where the numbers below the braces are those (in $10^{-5}$ units) obtained in the phenomenological toy model.

My conclusion is that the {\it moment analysis} approach described above may gradually lead to an accurate determination of $a_{\mu}^{\rm HVP}$, providing at the same time many tests of lattice QCD evaluations to be confronted with phenomenological determinations using experimental data.

\newpage
\begin{center}
{\normalsize\bf Acknowledgments.}
\end{center}

I am very grateful to Laurent Lellouch for many helpful discussions and encouragement.  I also thank the support of the OCEVU Labex (ANR-11-LABX-0060) and the A*MIDEX project (ANR-11-IDEX-0001-02) funded by the "Investissements d'Avenir" French government program managed by the ANR. 
 
\vspace*{0.25cm}


\end{document}